\documentclass{optica-article}

\journal{opticajournal} 

\articletype{Research Article}

\usepackage{lineno}

\usepackage{graphicx}
\usepackage{indentfirst}
\usepackage{braket}
\usepackage{float}
\usepackage{amsmath}
\usepackage{physics}

\usepackage{amssymb}
\usepackage{CJK}
\usepackage{esint}
\usepackage{color}
\usepackage[T1]{fontenc}
\usepackage{subfigure}
\usepackage{amsfonts}
\usepackage{footmisc}
\usepackage{scrextend}
\usepackage{multirow}
\usepackage[outdir=./]{epstopdf}
\usepackage{svg}
\usepackage{algorithm}
\usepackage{algpseudocode}
\usepackage{soul}
\usepackage{mathtools}
\usepackage[english]{babel}
\usepackage{url}
\usepackage{comment}
\usepackage{array}
\usepackage{subfiles}
\usepackage{tikz}
\usetikzlibrary{shapes,arrows}
\usepackage{mathrsfs}
\usepackage[mathscr]{eucal}

\usepackage{cleveref}
\usepackage{stmaryrd}
\usepackage{cite}

\definecolor{darkblue}{rgb}{0,0,0.5}
\hypersetup{
colorlinks=true,
linkcolor=black,
filecolor=blue,
citecolor=darkblue,  
urlcolor=black,
}

\usepackage{trimclip}

\makeatletter
\DeclareRobustCommand{\shortto}{%
  \mathrel{\mathpalette\short@to\relax}%
}

\newcommand{\short@to}[2]{%
  \mkern2mu
  \clipbox{{.5\width} 0 0 0}{$\m@th#1\vphantom{+}{\shortrightarrow}$}%
  }
\makeatother

\usepackage{pict2e,picture,graphicx}

\makeatletter
\DeclareRobustCommand{\Arrow}[1][]{%
\check@mathfonts
\if\relax\detokenize{#1}\relax
\settowidth{\dimen@}{$\m@th\rightarrow$}%
\else
\setlength{\dimen@}{#1}%
\fi
\sbox\z@{\usefont{U}{lasy}{m}{n}\symbol{41}}%
\begin{picture}(\dimen@,\ht\z@)
\roundcap
\put(\dimexpr\dimen@-.7\wd\z@,0){\usebox\z@}
\put(0,\fontdimen22\textfont2){\line(1,0){\dimen@}}
\end{picture}%
}
\makeatother
\newcommand{\veryshortrightarrow}{\hspace{.2mm}\scalebox{.8}{\Arrow[.1cm]}\hspace{.2mm}}

\def\be{\begin{equation}}
\def\ee{\end{equation}}
\def\ba{\begin{eqnarray}}
\def\ea{\end{eqnarray}}
\def\bal{\begin{equation}\begin{aligned}}
\def\eal{\end{aligned}\end{equation}}
\def\o{\overline}

\def\bp{\begin{pmatrix}}
\def\ep{\end{pmatrix}}

\usepackage{bm}

\newcommand{\calE}{{\cal E}}

\usepackage[normalem]{ulem}

\begin{document}

\title{Quantum illumination via frequency-mode-based correlation-to-displacement conversion}

\author{Xin Chen,\authormark{1,*} and Zhibin, Ye\authormark{1}}

\address{\authormark{1}The College of Electrical and Information Engineering, Quzhou University, Quzhou 324000, China.}

\email{\authormark{*}chenxin@qzc.edu.cn} 


\begin{abstract*} 
Quantum illumination leverages entanglement to surpass classical target detection, even in high-noise environments. Remarkably, its quantum advantage persists despite entanglement degradation caused by environmental decoherence. A central challenge lies in designing optimal receivers to exploit this advantage, with the correlation-to-displacement conversion module emerging as a promising candidate. However, the practical implementation of the conversion module faces technical hurdles, primarily due to the vast number of modes involved. In this work, we address these challenges by proposing a frequency-mode entangled source with matched photon numbers, a heterodyne detection scheme for the returned signals across vast modes, and a cavity-enhanced quantum pulse gate for programmable mode processing. This integrated framework paves the way for the realization of practical quantum illumination systems.

\end{abstract*}

\section{Introduction}
Quantum entanglement offers significant advantages for sensing \cite{Wineland_1992,Dowling_1998,giovannetti2006,giovannetti2011advances,Afek_2010,sidhu2020geometric,lawrie2019quantum,toth2014quantum,zhang2021dqs,Brady2023,Xia2023} and communication \cite{gisin2007quantum,wehner2018quantum,wilde2013quantum,kimble2008quantum}, though these benefits are fragile in the presence of environmental noise. As quantum decoherence increases, the entanglement that enables these advantages is rapidly destroyed. Quantum Illumination (QI), an entanglement-enhanced sensing scheme for target detection, has shown surprising resilience to environmental noise and loss~\cite{shapiro2020quantum,zhuang2021quantum,Lloyd2008,zhang2015,Barzanjeh_2015}. Theoretical predictions suggest a 6-dB error-exponent advantage over classical illumination (CI) in QI \cite{tan2008quantum}. However, practical receivers—such as the optical parametric amplifier (OPA) receiver \cite{Guha2009} demonstrated experimentally \cite{zhang2015}—can only achieve a 3-dB advantage in error exponent. Achieving the optimal receiver performance requires near-unit-efficiency sum-frequency generation (SFG) at the single-photon level~\cite{Zhuang2017}, which remains a significant experimental challenge.

To fully exploit the advantages of entanglement, a recently proposed optimal receiver—the correlation-to-displacement (C$\veryshortrightarrow$D) conversion module—is a promising approach \cite{shi2022}. This technique involves sending a significant number of modes to the target and performing separate heterodyne detection on the returned modes. Based on the classical measurement outcomes, the conditional idler modes are linearly combined into either a displaced thermal state (when the target is present) or a thermal state (when the target is absent). This transforms the detection problem into one of discriminating between these two states. Although the C$\veryshortrightarrow$D receiver achieves the optimal error probability for QI, several technical challenges must be overcome for practical implementation. These challenges stem primarily from the large number of modes involved—typically $10^{5\sim 7}$. Key issues include generating a source that produces numerous correlated pairs, each sharing the same mean photon number; performing independent heterodyne detection measurements across a vast number of return modes; and implementing a passive linear transformation to effectively combine the corresponding conditional idler modes into a single mode.

In this letter, we propose that entangled sources for the C$\veryshortrightarrow$D conversion module can be generated using spontaneous parametric down-conversion (SPDC) pumped by a single-frequency laser~\cite{ou2007multi}. By selecting the appropriate frequency band, it is feasible to generate 
$10^{5\sim7}$ pairs of correlated frequency modes with matching mean photon numbers. The separate heterodyne detection \cite{wiseman2010quantum} of the returned modes can be performed through continuous time-domain measurements. Additionally, we design a quantum pulse gate (QPG) \cite{Eckstein:11,PhysRevA.90.030302,Reddy:14,Reddy:18} enhanced by a cavity to select the desired temporal mode (TM) \cite{Raymer_2020} of the idler field, enabling the required passive linear transformation. This approach provides a viable path for realizing quantum illumination and could also be applied to quantum communication and other quantum-enhanced sensing tasks, such as phase estimation \cite{shi2022}. 

\begin{figure}[t]
    \centering
    \includegraphics[width=0.8\linewidth]{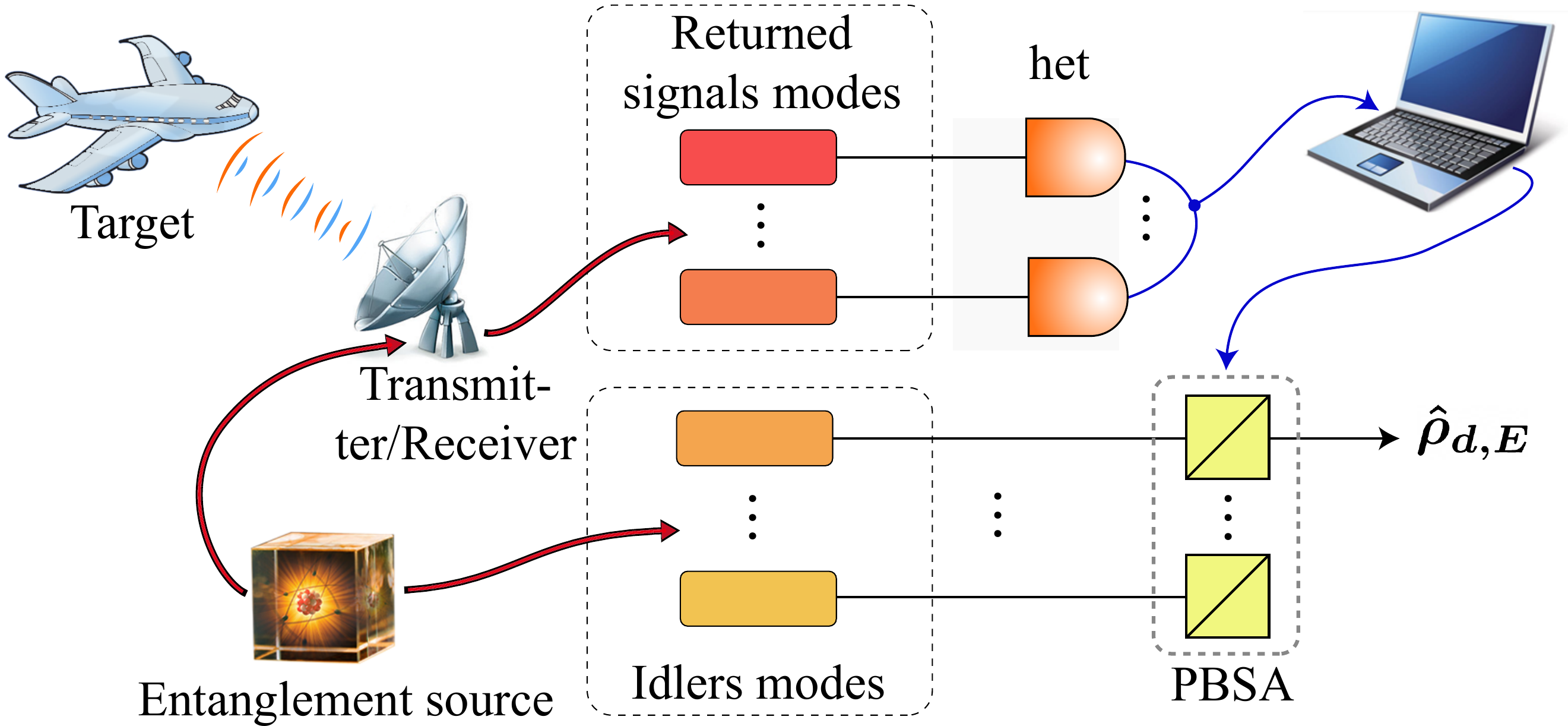}
    \caption{Schematic of quantum illumination featuring a practical receiver based on correlation-to-displacement conversion, in presence of noise
and loss. ‘het’: Heterodyne detection. ‘PBSA’: Programmable beam-splitter array.}
    \label{fig:QI}
\end{figure}
\section{Overall protocol of the C$\veryshortrightarrow$D conversion module}
As illustrated in Fig. \ref{fig:QI}, in a QI target detection scenario, a probe signal entangled with an ancilla is transmitted toward the target in a highly lossy and noisy environment. The return signal is jointly measured with the locally stored ancilla to infer the presence or absence of the target.

The C$\veryshortrightarrow$D conversion module utilizes $M$ signal-idler pairs $\{\hat{a}_{S_m},\hat{a}_{I_m}\}_{m=1}^M$, where each pair is in a two-mode squeezed-vacuum (TMSV) state with a mean photon number $N_S$.  In the ideal case of a known phase and a fixed target reflectivity, signals traverse a phase-shifted thermal-loss channel $\Phi_{\kappa,\theta}$ with transmissivity $\kappa$ (target reflectivity) and phase shift $\theta$. The return mode is given by:
\be
\hat{a}_{R_m}={\rm e}^{i\theta}\sqrt{\kappa}\hat{a}_{S_m}+\sqrt{1-\kappa}\hat{a}_{B_m},
\ee
where $\hat{a}_{B_m}$ represents a thermal state with a mean photon number $N_B / (1 - \kappa)$. When the target is absent ($\kappa = 0$), the channel reduces to $\Phi_{0,0}$ dominated by thermal noise. A heterodyne measurement is then performed on each $\hat{a}_{R_m}$, yielding the complex outcomes $\boldsymbol{\alpha}=\{\alpha_1,\cdots,\alpha_M\}^T$, where each $\alpha_m$ follows a circularly symmetric complex Gaussian distribution with variance $(N_B+\kappa N_S+1)/2$.
Conditioned on $\boldsymbol{\alpha}$, each idler collapses to a displaced thermal state $\hat{\rho}_{d_m,E}$ with complex displacement $d_m=\mu_{\kappa}{\rm e}^{i\theta}\boldsymbol{\alpha}_m^{*}$ and thermal noise $E=N_S(N_B+\kappa-1)/(N_B+\kappa N_S+1)$, where $\mu_{\kappa}=\sqrt{\kappa N_S(N_S+1)}/(\kappa N_S+N_B+1)$. Applying the passive linear optical transform described in Ref. \cite{shi2022}, a beamsplitter array with weights $w_m = \alpha_m / |\boldsymbol{\alpha}|$ combines idlers into a single-mode displaced thermal state $\hat{\rho}_{d, E}$ with total displacement $d=\sum w_m d_m=\mu_{\kappa}{\rm e}^{i\theta}|\boldsymbol{\alpha}|$. This yields the error probability performance limit
\be
P_{\rm C\veryshortrightarrow D}=\int_0^\infty dx P^{M}(x;\xi_{\kappa}) P_{\rm H}(\hat{\rho}_{0,N_S},\hat{\rho}_{\sqrt{x},E}),
\label{PCD}
\ee
where $x=\mu_{\kappa}^2|\boldsymbol{\alpha}|^2$ follows a $\chi^2$ distribution parameterized by $\xi_{\kappa}$
\be
P^{M}(x;\xi_{\kappa})=\frac{x^{M-1}{\rm e}^{-x/(2\xi_{\kappa})}}{(2\xi_{\kappa})^{M}\Gamma(M)},
\ee
with $\xi_{\kappa}=\kappa N_S(N_S+1)/2(\kappa N_S+N_B+1)$ and $\Gamma(M)=(M-1)!$. The Helstrom limit~\cite{Helstrom_1967,Helstrom1969,Helstrom_1976} for discriminating vacuum $\hat{\rho}_{0,N_S}$ (target absent) from $\hat{\rho}_{\sqrt{x},E}$ (target present) with equal prior probability is: 
\be
P_{\rm H}(\hat{\rho}_{0,N_S},\hat{\rho}_{\sqrt{x},E})=\frac{1}{2}\big(1-\frac{1}{2}{\rm Tr}[|\hat{\rho}_{0,N_S}-\hat{\rho}_{\sqrt{x},E}|]\big).
\label{Hellim}
\ee
\begin{figure*}
    \centering
    \includegraphics[width=0.8\linewidth]{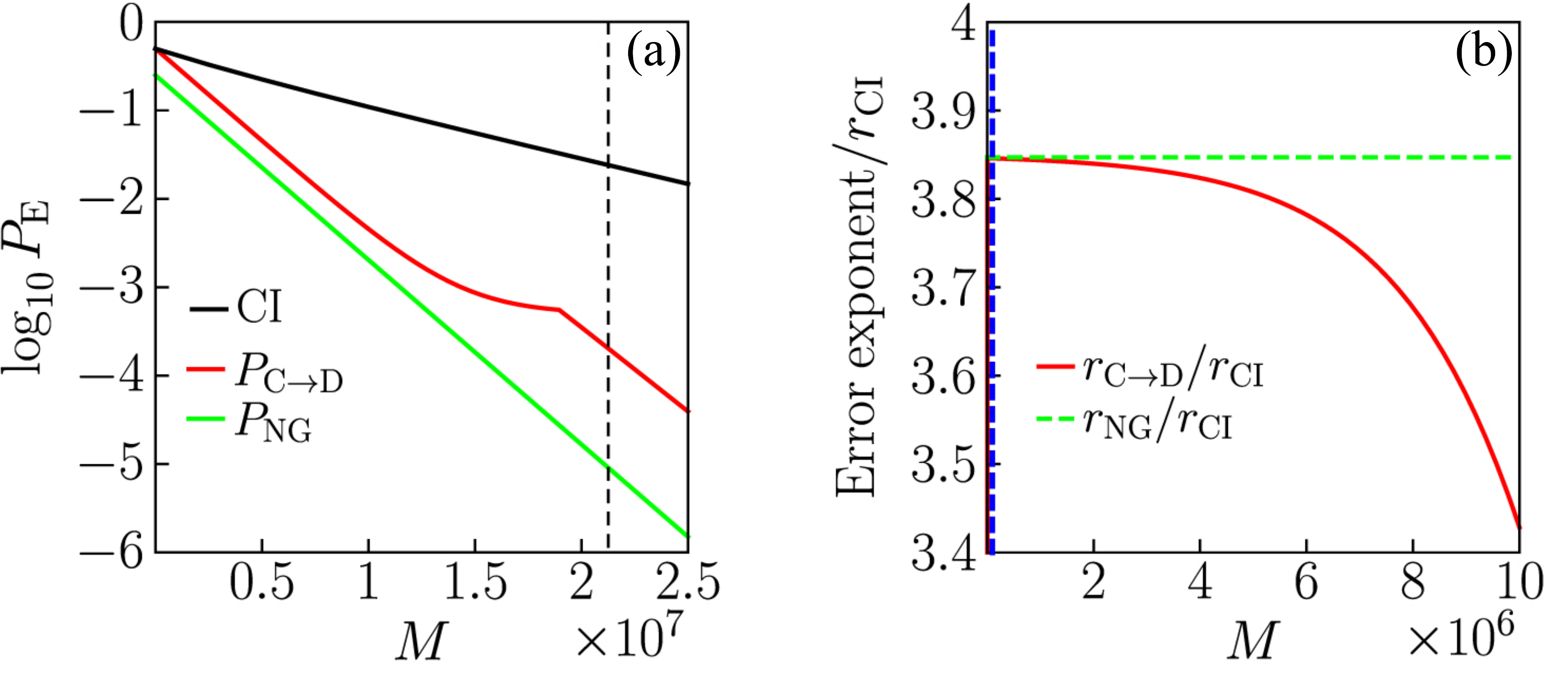}
   \caption{Error performance for a fixed-phase and known-reflectivity target with parameters: $N_{{\rm S}}=0.001$, $N_{\rm E}=20$ and $\kappa=0.01$. (a) Error probability as a function of the number of copies $M$. The red line indicates the error probability limit of the C$\veryshortrightarrow$D conversion, calculated according to \cite{PhysRevA.107.062405}. The discontinuity arises due to the optimal photon count decision threshold transitioning from $0$ to $1$. The green line represents the NG lower bound, while the black line indicates the optimal error probability for CI, as derived in \cite{PhysRevA.107.062405}. The vertical black dashed line marks the mode number where the optimal photon number threshold is $1$, approximated by $-W_{-1}(-N_S/{\rm e})/2\xi_{\kappa}$.
    (b) Error exponent ratio as a function of the number of copies $M$, where the error exponent for C$\veryshortrightarrow$D conversion is given by $r_{\rm C\veryshortrightarrow D}=-d{\rm ln}P_{\rm C\veryshortrightarrow D}/dM$, the error exponent for the NG bound by $r_{\rm NG}=-d{\rm ln}P_{\rm NG}/dM$, and the CI error exponent is assumed $r_{\rm CI}=\kappa N_S/4N_B$ \cite{Guha2009}. The vertical green dashed line indicates the mode count at $1\times 10^5$.}
    \label{fig2}
\end{figure*}

In the limit of low brightness $N_S\ll 1$, low transmissivity $\kappa\ll 1$ and the mode number $M\ll-W_{-1}(-N_S/{\rm e})/2\xi_{\kappa}$ (as shown in Fig. \ref{fig2}(a)), where $W_{-1}$ is the Lambert $W$ function (see Appendix \ref{suppnote}), $\hat{\rho}_{\sqrt{x},E}$ approximates a coherent state, and $\hat{\rho}_{0,N_S}$ approaches vacuum. The Helstrom limit becomes $P_{\rm H}\simeq {\rm e}^{-x}/4$, which—combined with Eq. (\ref{PCD})—gives the error exponent $r_{\rm C\veryshortrightarrow D}=2\xi_{\kappa}\simeq \kappa N_S/N_B$. In comparison, the optimal CI, using the coherent-state transmitter with mean photon number $N_S$ for each mode, has the error exponent $r_{\rm CI}=\kappa N_S/4N_B$ \cite{Guha2009}. The 6-dB advantage for QI over CI is achieved. We confirm this optimality in Fig. \ref{fig2}(a), where a close agreement is observed between $P_{\rm C\veryshortrightarrow D}$ (red) and the Nair-Gu (NG) lower bound $P_{\rm NG}$ (green) as given in Eq. (\ref{NGb}) (see Appendix \ref{NGbs})~\cite{nair2020fundamental}. 

In more practical scenarios with fading targets, random-phase noise and fluctuating transmissivity render the quantum channel non-Gaussian. The effective channel becomes
\cite{PhysRevA.107.062405}:
\be
\Bar{\Phi}=\int d\theta d\kappa P_{\Theta}(\theta)P_K(\kappa)\Phi_{\kappa,\theta},
\ee
with a time-independent random transmissivity, distributed according to $P_K(\cdot)$, and a random phase shift distributed according to $P_{\Theta}(\cdot)$. The heterodyne detection of the return modes projects idlers into mixed states diagonal in the Fock basis. Photon counting optimizes post-processing in this regime~\cite{PhysRevA.107.062405}, preserving a reduced (but nonzero) quantum advantage over CI.
\section{Signal source}
The C$\veryshortrightarrow$D conversion module requires a source of $M$ signal-idler pairs in a TMSV state with the same mean photon number, which can be achieved by SPDC pumped by a single-frequency mode laser. We show that it provides a series of one-to-one entangled fields in the frequency mode as follows (Note that the term `single mode' is only an approximation, as it is defined over a finite time duration, resulting in a limited bandwidth): the single mode pump field is described by $E_{\text{pump}}(t)=E_P{\rm e}^{i\omega_P t}$ with the complex amplitude $E_P$ and angular frequency $\omega_P$. Consider a transmission that lasts $T$ seconds, which is also the measurement duration for detection. It is reasonable to decompose the corresponding output fields of SPDC as 
$
\hat{a}_O(t)=(1/\sqrt{T})\sum_{n=-\infty}^{\infty} \hat{a}_{On} {\rm e}^{-i(\omega_n+\omega_O) t},
\label{at}
$
where $O\in\{S,I\}$ denotes the signal and idler fields, respectively,
$\omega_n=2\pi n/T$ and $\hat{a}_{On}=(1/\sqrt{T})\int_{-T/2}^{T/2}\hat{a}_O(t){\rm e}^{i(\omega_n+\omega_O) t}dt$ for $t\in(-T/2,T/2)$. $\hat{a}_{On}$ represents the single frequency mode with the angular frequency $\omega_n+\omega_O$. The carrier frequencies satisfy the energy conservation $\omega_P=\omega_S+\omega_I$. The operators mentioned above satisfy the commutation relations: $[\hat{a}_{Oi},\hat{a}^{\dagger}_{Oj}]=\delta_{i,j}$ and $[\hat{a}_O(t),\hat{a}_O^{\dagger}(t')]=\delta(t-t')$. Here we neglect the position variables for conciseness. Thus, the input-output relation for the SPDC process can be written as (see Appendix \ref{appspdc}) 
\be
\hat{a}_{S_n}=G\hat{a}^{0}_{S_n}+g\hat{a}^{0\dagger}_{I_{-n}},
\label{inpout1}
\ee
\be
\hat{a}_{I_n}=G\hat{a}^{0}_{I_n}+g\hat{a}^{0\dagger}_{S_{-n}},
\label{inpout2}
\ee
where the input fields $\hat{a}^{0}_{S_m}$ and $\hat{a}^{0}_{I_m}$ are in vacuum states. The constant Bogoliubov coefficients $G$ and $g$ arise from the phase-matching approximation for narrow frequency spans (with bandwidth $\Omega$) around the carrier frequencies, and they satisfy the relation $|G|^2-|g|^2=1$.  Eqs. (\ref{inpout1}) and (\ref{inpout2}) reveal $M=2l+1$ entangled signal-idler pairs $\{\hat{a}_{S_n},\hat{a}_{I_{-n}}\}_{n=-l}^{l}$ in TMSV states with mean photon number $N_S=|g|^2$. The mode count $l\sim\Omega T/4\pi$ is determined by the bandwidth $\Omega$ and the frequency separation between modes $2\pi/T$.
\section{Heterodyne detection}
The C$\veryshortrightarrow$D conversion module requires independent heterodyne detection on each return mode $\hat{a}_{R_n}$. This is implemented via continuous time-domain heterodyne measurement of $\hat{a}_R(t)$. The outcome for mode $n$ is $\nu\beta_{n}=(1/\sqrt{T})\int_{-T/2}^{T/2}\nu\beta(t){\rm e}^{i(\omega_n+\omega_S) t}dt$, where $\nu\beta_{n}$ and $\nu\beta(t)$ represent the measurement results for the operators $\hat{a}_{R_n}$ and $\hat{a}_R(t)$, respectively. The constant $\nu$ is a normalization factor chosen such that $\sum_n |\beta_n|^2=\int^{T/2}_{-T/2}|\beta(t)|^2dt=1$. This approach is analogous to the methodology described in \cite{PhysRevA.62.022309,PhysRevLett.99.110503}, where continuous quadrature measurements in the time domain provide classical information for the multi-mode teleportation. 
\begin{figure}[t]
    \centering
    \includegraphics[width=0.5\linewidth]{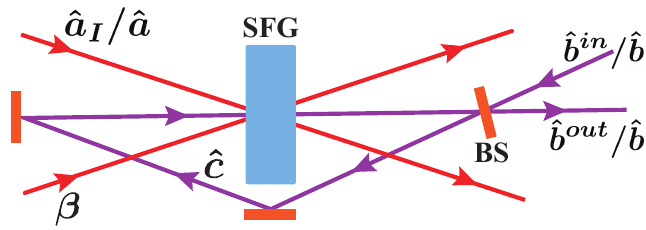}
    \caption{Schematic illustration of the cavity-enhanced quantum pulse gate proposal.}
    \label{fig:qpg}
\end{figure}
\section{Linear transform}
After heterodyne detecting each return mode individually, a passive linear transformation combines the corresponding conditional idler modes into a single mode—in our case, the TM $\hat{A}_0=\sum_{n} \beta_{-n}\hat{a}_{I_n}=\int_{-T/2}^{T/2}\beta(t)\hat{a}_I(t)dt$—for post process. $\{\beta_n\}$ and $\beta(t)$ are the spectral amplitude and temporal field of the TM, respectively. This mode combination is achieved using a specially designed QPG.
\subsection{Quantum pulse gate transform}
QPG is a cutting-edge technique for selectively sorting photonic TMs. Conceptually, a QPG functions as a quantum mechanical beamsplitter selectively reflects a single TM from a broadband multimode signal, transforming it into a different mode while leaving the other TMs transparent. Previous work by A. Eckstein et al. introduced a QPG based on nonlinear SFG \cite{Eckstein:11}. Notably, multi-stage SFG interferometry has been employed to enhance mode selectivity, achieving near-unit efficiency \cite{PhysRevA.90.030302}. However, current implementations still face challenges in achieving precise control over the desired TM profile and in resolving its frequency components. To overcome these limitations and meet the requirements of our application, we propose a novel QPG based on nonlinear SFG, along with an enhancement cavity, as illustrated in Fig. \ref{fig:qpg}. 
The total Hamiltonian for the SFG process enhanced by a cavity is given by:
\bal
\hat{H}/\hbar&=\omega_P\hat{c}^{\dagger}(t)\hat{c}(t)+\sum_n [(\omega_n+\omega_I)\hat{a}^{\dagger}_n(t)\hat{a}_n(t)+(\omega_n+\omega_P)\hat{b}^{\dagger}_n(t)\hat{b}_n(t)]\\&+i\sqrt{\gamma} [\hat{b}^{\dagger}(t)\hat{c}(t)-\hat{c}^{\dagger}(t)\hat{b}(t)]-i \eta [\hat{a}(t)\hat{c}^{\dagger}(t)\beta(t)-\hat{a}^{\dagger}(t)\hat{c}(t)\beta^{*}(t)],
\label{Hamitonianall}
\eal
where $\hat{a}(t)$, $\hat{b}(t)$, and $\hat{c}(t)$ are the bosonic operators for the signal mode, cavity mode, and the external field mode coupled to the cavity, respectively. A strong, non-depleting pump beam is generated by a programmable wave shaper, with its amplitude proportional to $\beta(t)$. 
 The modes $\hat{a}(t)$ and $\hat{b}(t)$ can be decomposed into their respective frequency components $\hat{a}(t)=(1/\sqrt{T})\sum_n\hat{a}_{n}(t)$ and $\hat{b}(t)=(1/\sqrt{T})\sum_n\hat{b}_{n}(t)$. The third term in the Hamiltonian represents the coupling between the cavity mode and the external field. The parameter $\gamma$ is the cavity decay rate, related to the coupling coefficient $K = \sqrt{\gamma / T}$ under the Markovian approximation. It corresponds to the cavity linewidth in the absence of internal losses. The fourth term describes the SFG interaction, with $\eta$ denoting the nonlinear coupling strength, proportional to the second-order nonlinear susceptibility $\chi^{(2)}$ and the amplitude of the pump field. The conditional idler modes $\hat{a}_{I_n} = \hat{a}_n(-T/2) {\rm e}^{-i \omega_n T/2}$ from the signal source are injected into the signal channel of the SFG as input modes. The idler channel is enhanced by the cavity at the fundamental frequency $\omega_P$, with the cavity mode $\hat{c}(t)=(1/\sqrt{T})\sum_n \hat{c}_{n}{\rm e}^{-i(\omega_n+\omega_P) t}$, where $\hat{c}_n$ 
is the corresponding Fourier coefficient for the frequency component. Adjacent resonances can be filtered out using intracavity filters, such as a grating structure \cite{Bai2021}. Notably, the carrier frequency of the idler channel does not need to match the signal source;  we set it to $\omega_P$ for convenience. The input-output relation for the system is derived as (see Appendix \ref{QPGSec}):
\be
\hat{b}^{out}_{n}=\frac{\frac{\eta\sqrt{\gamma}}{\sqrt{T}}}{i\omega_n-\frac{\gamma}{2}-\frac{\eta^2}{2T}}\hat{A}_n+\frac{i\omega_n+\frac{\gamma}{2}-\frac{\eta^2}{2T}}{i\omega_n-\frac{\gamma}{2}-\frac{\eta^2}{2T}}\hat{b}^{in}_{n}
\ee
where the TM $\hat{A}_{n}=\sum_l\beta_{n-l}\hat{a}_{I_l}$. Here, $\hat{b}^{in}_n=\hat{b}_{n}(-T/2){\rm e}^{-i\omega_n T/2}$ and $\hat{b}^{out}_n=\hat{b}_{n}(T/2){\rm e}^{i\omega_n T/2}$  represent the input and output field modes of the cavity, respectively, sharing the same carrier frequency $\omega_P$.  Under the condition $\eta=\sqrt{\gamma T}\ll\sqrt{2\pi}$, the output modes are: $\hat{b}^{out}_0=-\hat{A}_0$ and $\hat{b}^{out}_n=\hat{b}^{in}_n$ for $n\neq 0$.
This process fully converts the mode $\hat{A}_0$, while transmitting all other orthogonal modes in the input channel unchanged. The output field is then ready for post-processing, such as photon counting.
\subsection{Homodyne detection}
In certain scenarios, such as for an ideal target in the limits of low brightness $N_S\ll1$ and low reflection $\kappa\ll 1$, the detection decision is made by distinguishing between an approximate coherent state and an approximate vacuum state, as discussed in the previous section. A simple homodyne receiver provides an near-optimal measurement strategy that reduces technological complexity \cite{PhysRevLett.101.210501}. Homodyne detection inherently selects a TM through spectral overlap with a broadband local oscillator (LO), thus bypassing the need for the QPG transformation step. The process unfolds as follows: a LO field pulse is prepared in the form of
$
\varepsilon_{LO}(t)=(|\calE|{\rm e}^{i\theta}/\sqrt{T})\sum_n \beta_{-n}^*{\rm e}^{-i(\omega_n+\omega_I) t},
$
where the amplitude of the LO field is strong: $|\calE|\gg 1$. Notably, the carrier frequency of the LO field $\omega_I$ differs from the pump field's carrier frequency $\omega_S$ used in the QPG. The detected quadrature is then \cite{ou2017quantum}
\bal
\int_{-T/2}^{T/2}[\varepsilon_{LO}^*(t)\hat{a}_I(t)+h.c]dt&=\int_{-T/2}^{T/2}\big[\frac{|\calE|{\rm e}^{-i\theta}}{T}\sum_{n,l} \beta_{-n}\hat{a}_{I_l} {\rm e}^{-i(\omega_l-\omega_n) t}+h.c.\big]dt\\&=|\calE|\hat{Q}(\theta),
\label{homo}
\eal
where $\hat{Q}(\theta)={\rm e}^{-i\theta}\hat{A}_{0}+h.c.$. In the second line of Eq. (\ref{homo}), we have used the identity $(1/T)\int_{-T/2}^{T/2}{\rm e}^{-i(\omega_{l}-\omega_{n}) t}=\delta_{l,n}$. The discrimination decision is then made by comparing the outcome of $\hat{Q}(\theta)$ with a predefined threshold.
\section{Discussion}
Finally, we specify realistic parameters to evaluate the feasibility of experimentally implementing this scheme. For the signal source, assume the frequency regime $\Omega$ has a bandwidth of approximately 10 GHz, with modes that share the same mean photon number (see Appendix \ref{appspdc}). The single-frequency pump laser used for SPDC has a linewidth on the order of 1 MHz \cite{Bai2021}. We set the measurement time duration $T$ equal to the coherent time of the pump laser, which corresponds to a frequency interval of $2\pi/T \sim 1$ MHz. This leads to an estimated mode number of $\Omega T / (2\pi) \sim 10^5$.

For the QPG transformation, the cavity linewidth must be significantly smaller than the frequency interval, i.e., $\gamma \ll 2\pi/T$. Thus, a cavity with a linewidth of $\gamma \sim 10$ kHz is suitable~\cite{Bai2021,8103793}. Assuming the mean photon number of signal $N_S=0.001$, the reflectivity of the target $\kappa=0.01$ and the scaled environment noise photon number $N_B=20$, this configuration yields a 5.85-dB error-exponent
advantage over CI
, as predicted by the theory for an ideal target with fixed phase shift and known reflection. It is marked by the vertical green dashed line in the Fig. \ref{fig2}(b) which plots the error exponent ratio $r_{\rm C\veryshortrightarrow D}/r_{\rm CI}$ for a broad range of mode numbers $M$.

If further optimization is desired, the mode number can be increased by 2-3 orders of magnitude by using a narrower-linewidth pump laser (around 1 kHz) and a cavity with a linewidth of approximately 10 Hz for SPDC and QPG \cite{Bai2021,Fan:12,Xu:13,Lewoczko-Adamczyk:15}, respectively. However, this would come at the expense of increased technical complexity.
\section{Conclusion}
We address several key challenges in the practical implementation of the recently proposed optimal receiver, the C$\veryshortrightarrow$D conversion module, which fully leverages the 6-dB error-exponent advantage over optimal classical illumination in low-signal-brightness scenarios. We also assess the feasibility of these solutions within the constraints of current experimental techniques. Furthermore, our approach not only provides a promising pathway for realizing quantum illumination but also has the potential to be extended to quantum communication and other quantum-enhanced sensing applications, such as phase estimation.
\appendix
\section{C$\veryshortrightarrow$D conversion module supplementary notes}
\label{suppnote}
In this section, we analyze the error probability performance for an ideal target, characterized by fixed, known reflectivity and phase shift, in the limit of low brightness $N_S\ll1$ and low reflectivity $\kappa\ll1$. Initially, we demonstrate that the error probability for both the ideal and random-phase targets (with fixed, known reflectivity and uniformly distributed phase shift) is identical. Subsequently, we focus on analyzing the error probability performance for the ideal target by leveraging the results for the random-phase target.

According to \cite{PhysRevA.107.062405}, for a random-phase target, the density matrix element of the outcome state after a passive linear transformation in the number basis is given by
\(
\hat{\rho}^U_{m,n} = \delta_{m,n} \hat{\rho}_{n,n},
\)  
where \(\delta_{m,n}\) is the Kronecker delta, and \(\hat{\rho}\) is the density matrix for the ideal target. This implies that the diagonal elements of the density matrix remain unchanged across both scenarios. As a result, the Helstrom limit remains identical for both cases: $P_{\rm H}(\hat{\rho}_{0,N_S},\hat{\rho}^U_{\sqrt{x},E})=P_{\rm H}(\hat{\rho}_{0,N_S},\hat{\rho}_{\sqrt{x},E})$, as shown in Eq. (\ref{Hellim}), since the trace operation depends solely on the diagonal elements of the density matrices. Furthermore, the distribution of heterodyne detection results \(P^{M}(x; \mu_{\kappa})\) is independent of the phase shift. Therefore, the error probability performance is the same for both the ideal and random-phase targets, according to Eq.~(\ref{PCD}). 

Next, we analyze the performance for the ideal target by drawing on the results for the random-phase target. For the scenario of random-phase noise, photon counting is the optimal measurement strategy, since \(\hat{\rho}^U\) is diagonal in the number basis. The optimal photon number decision threshold is expressed as  
\(
N_{\text{opt}} = 2\mu_{\kappa}^2 (\kappa N_S + N_B + 1) M/\epsilon,
\)  
where \(\epsilon = -W_{-1}(-N_S / e)\) (see Eq.~31 in \cite{PhysRevA.107.062405}). In the low-brightness and low-reflectivity regime, the thermal state $\hat{\rho}_{0,N_S}$ can be approximated as a vacuum state, while the state $\hat{\rho}^U_{\sqrt{x},E}$ is approximated as the number-basis-diagonalized state $\hat{\rho}^U_{\sqrt{x},0}$, with the corresponding state $\hat{\rho}_{\sqrt{x},0}$ being a coherent state. This results in an error exponent that is four times (i.e., 6 dB) that of the CI, as stated in the main context for the ideal target. However, it is important to note that the optimal decision strategy for distinguishing between the state $\hat{\rho}^U_{\sqrt{x},0}$ and a vacuum state corresponds to a photon number decision threshold of \(N = 0\)—the Kennedy receiver—since the probability of detecting a photon count greater than zero is zero for a vacuum state (see Fig.~4a in \cite{PhysRevA.107.062405}). This implies that, for random-phase-target detection, in order to reach the approximated performance which fulfill the advantage of QI, an additional condition is required: the optimal photon number threshold must satisfy
\(
N_{\text{opt}} \ll 1
\). This leads to the following constraint on the mode number:
\(
M \ll \epsilon/2 \xi_{\kappa}
\), which also applies for the ideal target, as shown in Fig. \ref{fig2}(a).
\section{Nair-Gu bound}
\label{NGbs}
Nair and Gu derived a fundamental lower bound $P_{\rm NG}$ on the minimum error probability for QI target detection using arbitrary entangled resources~\cite{nair2020fundamental}. As this bound applies to idealized scenarios (e.g., noise-free operation with optimal receivers), it remains valid even in the presence of additional noise, thereby representing an ultimate quantum limit.

For $M$ entangled signal-idler probe pairs with mean signal photon number $N_{{\rm S}}$, the NG bound is given by
\be
P_{\rm C\veryshortrightarrow D} \geq P_{\rm NG}=\frac{1}{4}e^{-\beta M N_{{\rm S}}},
\label{NGb}
\ee
where $\beta=-\ln[1-\kappa/(N_E(1-\kappa)+1)]$.
\section{Spontaneous parametric downconversion}
\label{appspdc}
Consider a three-wave mixing SPDC process driven by a strong, non-depleting, single-frequency laser. The pump field is classically described as $E_{pump}=E_P{\rm e}^{-i(\omega_Pt-k_Pz)}$. To ensure optimal phase matching, we assume fixed polarizations for the signal, idler, and pump fields, allowing us to treat them as scalar fields.  The interaction Hamiltonian integrated over time is given by
\be
\hat{H}(z)=\int_{-T/2}^{T/2} \hat{H}(z,t)dt=i\hbar\zeta\int_{-T/2}^{T/2} \hat{a}_S(z,t)\hat{a}_I(z,t){\rm e}^{i(\omega_P t-k_Pz)}dt+h.c.,
\label{appH}
\ee
where the field operators $\hat{a}_O(z,t)=(1/\sqrt{T})\sum_{n=-\infty}^{\infty} \hat{a}_{O_n}(z) {\rm e}^{-i[(\omega_O+\omega_n) t-k_O(\omega_O+\omega_n)z]}$ and their Fourier components are $\hat{a}_{O_n}(z)=(1/\sqrt{T})\int_{-T/2}^{T/2}\hat{a}_O(z,t){\rm e}^{i[(\omega_O+\omega_n) t-k_O(\omega_O+\omega_n)z]}dt$. Here, \( O \in \{S, I\} \) represents the signal ($S$) and idler ($I$) fields, respectively, and $\zeta$  is the nonlinear coupling strength, proportional to the second-order susceptibility $\chi^{(2)}$ and the pump field amplitude. Solving the time integral yields
\be
\hat{H}(z)=i\hbar \zeta\sum_{n=-\infty}^{\infty}\hat{a}_{S_n}(z)\hat{a}_{I_{-n}}(z){\rm e}^{-i\Delta k_n z},
\label{appHf}
\ee
where $\Delta k_n=k_P-k_S(\omega_S+\omega_n)-k_I(\omega_I-\omega_n)$ is the so-called phase mismatch. We assume the downconverter is phase-matched at $\Delta k_0=0$ and that the frequency-sum condition $\omega_P=\omega_S+\omega_I$ is satisfied. Let the uniform-medium length be $L$, with the interaction starting at $z = 0$. Consider the small frequencies \(\omega_n\) such that \(\Delta k_n L \ll 1\), which correspond to frequencies confined within two distinct bands centered around the carrier frequencies \(\omega_S\) and \(\omega_I\), each with a bandwidth \(\Omega\). In this regime, the phase-matching approximation \(\Delta k_n z \approx 0\) applies, leading to
\be
\frac{\partial \hat{a}_{S_{n}}(z)}{\partial z}=\zeta \hat{a}_{I_{-n}}^{\dagger}(z),
\ee
\be
\frac{\partial \hat{a}_{I_{n}}(z)}{\partial z}=\zeta \hat{a}_{S_{-n}}^{\dagger}(z).
\ee
The solutions are
\be
\hat{a}_{S_{n}}(L)=G\hat{a}_{S_{n}}(0)+g\hat{a}_{I_{-n}}^{\dagger}(0),
\label{Sa}
\ee
\be
\hat{a}_{I_{n}}(L)=G\hat{a}_{I_{n}}(0)+g\hat{a}_{S_{-n}}^{\dagger}(0),
\label{Sb}
\ee
where $G={\rm cosh}(|\zeta|L)$ and $g=(\zeta/|\zeta|){\rm sinh}(|\zeta|L)$. In the main text, we use the shorthand $\hat{a}_{S_n}\equiv\hat{a}_{S_{n}}(L)$, $\hat{a}_{I_n}\equiv\hat{a}_{I_{n}}(L)$, $\hat{a}_{S_n}^0\equiv\hat{a}_{S_{n}}(0)$ and $\hat{a}_{I_n}^0\equiv\hat{a}_{I_{n}}(0)$. Eqs. (\ref{Sa}) and (\ref{Sb})  indicate that the output fields comprise $2l+1$ signal-idler pairs $\{\hat{a}_{S_n},\hat{a}_{I_{-n}}\}_{n=-l}^{l}$ in TMSV states with mean photon number $N_S=|g|^2$. The mode count $l\sim\Omega T/4\pi$ derives from the bandwidth of interest $\Omega$ and frequency resolution $2\pi/T$: The finite time
$T$ limits frequency discrimination to intervals of $2\pi/T$, yielding $\sim \Omega T/2\pi$ resolvable modes. Factor $1/2$ arises from symmetric pairing around carrier frequencies. The bandwidth of the down-converted fields (typically determined by $\Delta k_n L\sim1$) is usually $10^{12\sim 13}$ Hz \cite{ou2007multi}. Assuming $\Delta k_n \propto \omega_n^2$, as is typically the case, a condition of $\Delta k_n L=0.01$ results in a bandwidth $\Omega$ on the order of $10^{11\sim 12}$ Hz. 
\section{Cavity-enhanced quantum pulse gating}
\label{QPGSec}
In this section, we present a detailed analysis of the cavity-enhanced QPG results discussed in the main text.

The total Hamiltonian of SFG enhanced by a cavity reads
\bal
\hat{H}/\hbar&=\omega_P\hat{c}^{\dagger}(t)\hat{c}(t)+\sum_n [(\omega_n+\omega_I)\hat{a}^{\dagger}_n(t)\hat{a}_n(t)+(\omega_n+\omega_P)\hat{b}^{\dagger}_n(t)\hat{b}_n(t)]\\&+i\sqrt{\gamma} [\hat{b}^{\dagger}(t)\hat{c}(t)-\hat{c}^{\dagger}(t)\hat{b}(t)]-i \eta [\hat{a}(t)\hat{c}^{\dagger}(t)\beta(t)-\hat{a}^{\dagger}(t)\hat{c}(t)\beta^{*}(t)],
\label{Hamitonianall}
\eal
In the rotating frame defined by the transformations: $\hat{a}_n(t)\rightarrow\hat{a}_n(t){\rm e}^{-i\omega_I t}$, $\beta(t)\rightarrow\beta(t){\rm e}^{-i\omega_S t}$, $\hat{c}(t)\rightarrow\hat{c}(t){\rm e}^{-i\omega_P t}$ and $\hat{b}_n(t)\rightarrow\hat{b}_n(t){\rm e}^{-i\omega_P t}$, the Hamiltonian simplifies to
\bal
\hat{H}/\hbar&=\sum_n \omega_n[\hat{a}^{\dagger}_n(t)\hat{a}_n(t)+\hat{b}^{\dagger}_n(t)\hat{b}_n(t)]+i\sqrt{\gamma} [\hat{b}^{\dagger}(t)\hat{c}(t)\\&-\hat{c}^{\dagger}(t)\hat{b}(t)]-i \eta [\hat{a}(t)\hat{c}^{\dagger}(t)\beta(t)-\hat{a}^{\dagger}(t)\hat{c}(t)\beta^{*}(t)].
\label{Hamitonianallro}
\eal
From Eq.~\eqref{Hamitonianallro}, the Heisenberg equations of motion for $\hat{a}_n$, $\hat{b}_n$ and $\hat{c}$ are derived as
\be
\dot{\hat{a}}_{n}(t)=-i\omega_n\hat{a}_{n}(t)+\frac{\eta}{\sqrt{T}}\hat{c}(t)\beta^{*}(t)
\label{aeq}
\ee
\be 
\dot{\hat{b}}_n(t)=-i\omega_n\hat{b}_n(t)+\sqrt{\frac{\gamma}{T}}\hat{c}(t),
\label{b_sol}
\ee
\be
\dot{\hat{c}}(t)=-\eta\hat{a}(t)\beta(t)-\sqrt{\gamma}\hat{b}(t).
\label{ceq}
\ee
Solving Eqs.~\eqref{aeq} and \eqref{b_sol}, we obtain \cite{PhysRevA.31.3761}
\be
\hat{a}_{n}(t)={\rm e}^{-i\omega_n (t+\frac{T}{2})}\hat{a}_{n}(-\frac{T}{2})+\frac{\eta}{\sqrt{T}}\int_{-\frac{T}{2}}^t {\rm e}^{-i\omega_n(t-t')}\hat{c}(t')\beta^{*}(t')dt',
\ee
\be 
\hat{b}_n(t)={\rm e}^{-i\omega_n (t+\frac{T}{2})}\hat{b}_n(-\frac{T}{2})+\sqrt{\frac{\gamma}{T}}\int_{-\frac{T}{2}}^t{\rm e}^{-i\omega_n (t-t')}\hat{c}(t')dt',
\label{bt_sol}
\ee
yielding
\bal
\hat{a}(t)&=\frac{1}{\sqrt{T}}\sum_n\hat{a}_{n}(t)=\hat{a}_I(t)+\frac{\eta}{2}\hat{c}(t)\beta^{*}(t),
\label{amode}
\eal
\bal
\hat{b}(t)&=\frac{1}{\sqrt{T}}\sum_n\hat{b}_{n}(t)=\hat{b}^{in}(t)+\frac{\sqrt{\gamma}}{2}\hat{c}(t),
\label{bmode}
\eal
where $\hat{a}_I(t)=(1/\sqrt{T})\sum_n \hat{a}_{I_n} {\rm e}^{-i\omega_n t}$ and $\hat{b}^{in}(t)=(1/\sqrt{T})\sum_n \hat{b}^{in}_{n} {\rm e}^{-i\omega_n t}$
, with $\hat{a}_{I_n}=\hat{a}_{n}(-T/2){\rm e}^{-i\omega_n T/2}$ and $\hat{b}^{in}_n=\hat{b}_{n}(-T/2){\rm e}^{-i\omega_n T/2}$. In the derivation above, we have used the identity $
(1/T)\sum_n{\rm e}^{-i\omega_n (t-t')}=\delta(t-t')
$.
Substituting Eqs. (\ref{amode}) and (\ref{bmode}) into Eq. (\ref{ceq}), we obtain the Langevin equation for the cavity mode
\be
\dot{\hat{c}}(t)=-\frac{\gamma}{2}\hat{c}(t)-\eta\hat{a}_I(t)\beta(t)-\frac{\eta^2}{2}\hat{c}(t)|\beta(t)|^2-\sqrt{\gamma}\hat{b}^{in}(t),
\label{ceq2}
\ee
Solving Eq. (\ref{ceq2}) in the frequency domain, we obtain 
\be
(i\omega_n-\frac{\gamma}{2})\hat{c}_{n}-\frac{\eta^2}{2T}\sum_{m,l}\beta^{*}_{l+m-n}\beta_{m}\hat{c}_{l}-\frac{\eta}{\sqrt{T}}\hat{A}_n-\sqrt{\gamma}\hat{b}^{in}_{n}=0,
\label{ceqf}
\ee
Since $\{\nu\beta_{n}\}$  represents heterodyne measurement outcomes for each mode $\hat{a}_{R_n}$, these correspond to random samples from a complex Gaussian distribution. Therefore, in the limit of large sample sizes, we have
\be
\sum_{m}\beta^{*}_{l+m-n}\beta_{m}\approx
    \begin{cases}
      1, & \text{if}\ l=n \\
      0. & \text{otherwise}
    \end{cases}
\ee
 Thus, Eq.~\eqref{ceqf} simplifies to
\be
(i\omega_n-\frac{\gamma}{2}-\frac{\eta^2}{2T})\hat{c}_{n}-\frac{\eta}{\sqrt{T}}\hat{A}_n-\sqrt{\gamma}\hat{b}^{in}_{n}=0,
\ee
where the TM $\hat{A}_{n}=\sum_l\beta_{n-l}\hat{a}_{I_l}$. Solving for 
$\hat{c}_n$ yields:
\be
\hat{c}_{n}=\frac{\frac{\eta}{\sqrt{T}}}{i\omega_n-\frac{\gamma}{2}-\frac{\eta^2}{2T}}\hat{A}_n+\frac{\sqrt{\gamma}}{i\omega_n-\frac{\gamma}{2}-\frac{\eta^2}{2T}}\hat{b}^{in}_{n}.
\label{cn}
\ee
Defining the output field of the cavity as $\hat{b}^{out}(t)=(1/\sqrt{T})\sum_n \hat{b}^{out}_{n} {\rm e}^{-i\omega_n t}$ where $\hat{b}^{out}_n=\hat{b}_{n}(T/2){\rm e}^{i\omega_n T/2}$, we obtain the input-output relation for the cavity from Eq. (\ref{bt_sol})
\be
\hat{b}^{out}(t)-\hat{b}^{in}(t)=\sqrt{\gamma}\hat{c}(t).
\label{inout1}
\ee
By combining Eqs.~(\ref{cn}) and (\ref{inout1}), we obtain the input-output relation for
\be
\hat{b}^{out}_{n}=\frac{\frac{\eta\sqrt{\gamma}}{\sqrt{T}}}{i\omega_n-\frac{\gamma}{2}-\frac{\eta^2}{2T}}\hat{A}_n+\frac{i\omega_n+\frac{\gamma}{2}-\frac{\eta^2}{2T}}{i\omega_n-\frac{\gamma}{2}-\frac{\eta^2}{2T}}\hat{b}^{in}_{n}.
\ee
When $\frac{\gamma}{2}=\frac{\eta^2}{2T}$ and $\omega_1=\frac{2\pi}{T}\gg \frac{\gamma}{2}+\frac{\eta^2}{2T}=\frac{\eta^2}{2T}$ (namely, $\eta=\sqrt{\gamma T}\ll\sqrt{2\pi}$), we obtain
\be
\hat{b}^{out}_{n}=
    \begin{cases}
      \hat{A}_0, & \text{if}\ n=0 \\
      \hat{b}^{in}_{n}, & \text{otherwise}
    \end{cases}
\ee

\begin{backmatter}
\bmsection{Funding}
This work is supported by the Joint Fund of Zhejiang Provincial Natural Science Foundation of China (LQZSZ25F050001).

\bmsection{Disclosures}
The authors declare no conflicts of interest.

\bmsection{Data Availability Statement}
No data were generated or analyzed in the presented research.
\end{backmatter}

\bibliography{sample}






\end{document}